 \documentclass{aa}
\usepackage{amssymb}
\usepackage{hyperref}
\usepackage{times,graphicx}
\usepackage{ulem,widetext}

\DeclareMathAlphabet\mathbfcal{OMS}{cmsy}{b}{n}
\usepackage{mathtools}

\begin{document}

\title{Collisional and magnetic effects on the polarization of the solar oxygen infrared triplet}
\author{ M. Derouich\thanks{aldarwish@kau.edu.sa \& derouichmoncef@gmail.com} and S. Qutub\thanks{squtub@kau.edu.sa \& saleh.qutub@hotmail.com}}

\institute{ 
Astronomy and Space Science Department, Faculty of Science, King Abdulaziz University, P.O. Box 80203, Jeddah 21589, Saudi Arabia   }

\titlerunning{Depolarizing  effects on the solar polarization of O\,\textsc{i} triplet}
\authorrunning{Derouich \& Qutub}

\date{Accepted for publication}
\abstract
{The scattering polarization of the infrared (IR) triplet of neutral oxygen (O\,\textsc{i}) near 777\,nm  provides a powerful diagnostic of   solar atmospheric conditions. However, interpreting such polarization requires a rigorous treatment of isotropic depolarizing collisions between O\,\textsc{i} atoms and neutral hydrogen.}
{{We aim to investigate the combined effects of collisional and magnetic depolarization in shaping the alignment of  O\,\textsc{i} levels  (and thus the polarization of the O\,\textsc{i} IR triplet).}}
{We compute, for the first time, a comprehensive set of collisional depolarization and polarization transfer rates for the relevant O\,\textsc{i} energy levels. These rates are incorporated into a multi-level atomic model, and the statistical equilibrium equations (SEE) are solved to quantify the impact of collisions and magnetic fields on atomic alignment.}
{Our calculations indicate that elastic collisions with neutral hydrogen, together with the Hanle effect of turbulent magnetic fields stronger than about 20 G, efficiently suppress the bulk of the atomic alignment in deep photospheric conditions where hydrogen densities exceed $n_{\mathrm{H}} \sim 10^{16}$ cm$^{-3}$. In the chromosphere, however, the lower hydrogen density weakens collisional depolarization, allowing polarization to persist.}
{Our results are consistent with a chromospheric origin for the linear polarization signals of the O I IR triplet. Future studies should combine accurate non-LTE radiative transfer with reliable collisional rates in order to achieve fully consistent modeling.}

\maketitle

\section{Introduction }
In recent decades, there has been growing interest in finding reliable ways to detect, understand, and measure magnetic fields in the Sun’s photosphere and chromosphere. The advent of high-sensitivity polarimetry has enabled the detection of subtle polarization signals in numerous spectral lines, particularly those of the so-called Second Solar Spectrum (e.g. Stenflo 1994, Landi Degl'Innocenti \& Landolfi 2004, Trujillo Bueno 2001). These weak polarization signatures, often shaped by the Hanle effect, provide valuable insights into the  geometry and strength of magnetic fields in regions where the Zeeman effect alone is insufficient. However, interpreting scattering polarization signals observed in spectral lines in terms of solar magnetic fields remains notoriously challenging due to the complex interplay among atomic polarization, radiative transfer, magnetic fields, and collisional depolarization processes.

Since the 1990s, numerous studies have investigated various spectral lines and mechanisms for diagnosing solar magnetism. In this context, some attention has been given to the O\,\textsc{i} infrared (IR) triplet—a promising diagnostic target that  would benefit from  a more rigorous treatment, especially with regard to collisional effects. In fact, O\,\textsc{i} is one of the most abundant elements in the Sun, and the physical interpretation of its spectral lines can play a significant role in understanding the properties of the solar atmosphere (e.g., Asplund et al. 2021 and references therein).

High-precision observations of the scattering polarization of the O\,\textsc{i} IR triplet were reported by   Keller  \& Sheeley  (1999), Trujillo Bueno et al. (2001) and  Sheeley \& Keller (2003)     both on the disk close to the limb and off the limb.  These  observations revealed clear linear polarization signals, but they did not by themselves allow one to disentangle the relative contributions from photospheric and chromospheric layers.  

Theoretical investigations  of del Pino Alem\'an \& Trujillo Bueno (2015, 2017) addressed this problem by solving the non-LTE radiative transfer, including effects of elastic and inelastic collisions and magnetic fields via the Hanle effect. These studies showed that collisional depolarization in the deep photosphere is needed to reproduce the observed polarization patterns, and that the main contributions to the triplet’s polarization generally arise from the chromosphere. Motivated by these results, we focus here on quantifying with improved accuracy the role of elastic collisions between neutral hydrogen and O\,\textsc{i}, by providing precise depolarization and polarization-transfer rates and by illustrating the sensitivity of the atomic alignment of the O\,\textsc{i} levels to both hydrogen density and magnetic fields.

Concerning inelastic collisions between electrons and  O\,\textsc{i},   accurate quantum--mechanical calculations by Barklem (2007) provide an excitation rate coefficient of $\langle \sigma v \rangle = 1.02 \times 10^{-8}~\mathrm{cm^{3}\,s^{-1}}$ for the    $3s\,^{5}\!S \rightarrow 3p\,^{5}\!P$ transition (O\,\textsc{i}  IR triplet) at $T = 5000$~K. Using detailed balance, and assuming that the initial and final states have similar statistical weights, this corresponds to a de--excitation coefficient of $q_{\downarrow}  =4.13 \times 10^{-7}~\mathrm{cm^{3}\,s^{-1}}$. In the solar atmospheric regions relevant for this work, the electron density can reach up to  $N_e \approx 10^{12}~\mathrm{cm^{-3}}$ (e.g. Fontenla et al.\ 1993, Cox 2000), which implies a collisional rate of $C \approx 4 \times 10^{5}~\mathrm{s^{-1}}$. This is two orders of magnitude smaller than the radiative rate $A = 3.69 \times 10^{7}~\mathrm{s^{-1}}$ (NIST), yielding $C/A \approx 10^{-2}$, i.e., only at the percent level. 
Regarding inelastic  collision  couplings from \(3s\,^{5}\!S\) to the lower terms \(2p^{4}\,{}^{3}\!P\), \(2p^{4}\,{}^{1}\!D\), and \(2p^{4}\,{}^{1}\!S\); only the coupling to \(2p^{4}\,{}^{3}\!P\) is non-zero (see Barklem 2007), and its de--excitation rate is about 100 times smaller than that of \(3s\,^{5}\!S \rightarrow 3p\,^{5}\!P\), which is itself already negligible compared with the relevant radiative rates and with elastic O--H depolarization. Likewise, for the  forbidden \(3p\,^{5}\!P \rightarrow 3p\,^{3}\!P\) transition, the de--excitation rate is  smaller than for \(3s\,^{5}\!S \rightarrow 3p\,^{5}\!P\), and is therefore also negligible.  Therefore, in the photosphere the dominant collisional depolarizing mechanism is expected to arise from elastic collisions with neutral hydrogen atoms. However, pursuing a more quantitative, systematic investigation of e–O I collisions, using accurate quantum-mechanical inelastic rates, would be valuable in the future.

In this context,  our analysis focuses on the impact of elastic collisions, together with turbulent magnetic fields typical of the quiet Sun, when considering hydrogen densities typical of the photosphere and chromosphere.
We compute, for the first time, the full set of depolarization and polarization transfer rates for O\,\textsc{i} levels using a rigorous atomic approach, which allows us to make a substantial improvement through a detailed and quantitative treatment of collisional depolarization by elastic collisions with neutral hydrogen. We solve  the statistical equilibrium equations (SEE) using  a multi-level  atomic model,  following the formalism of Landi Degl’Innocenti  \& Landolfi (2004).  In our calculations we account for the depolarizing effect of isotropic collisions with neutral hydrogen atoms, as well as for the Hanle effect produced by turbulent magnetic fields, understood here as magnetic fields of constant strength but with orientations that vary randomly at sub-resolution scales, with an isotropic distribution (see, e.g. Landi Degl'Innocenti \& Landolfi 2004). 
 For typical quiet-Sun magnetic field strengths,   we assess the impact of collisional effects by comparing the atomic alignment—responsible for linear polarization—obtained with and without the inclusion of collisions. This analysis enables a   characterization  of the combined effects of magnetic fields and collisions on atomic alignment, and helps identify the physical conditions under which polarization signals can persist or be suppressed.

In the following, Sect. 2 describes the  generation  of atomic polarization in the O\,\textsc{i} levels involved in the modeling of the IR triplet. In Sect. 3, we present the formalism used to account for magnetic depolarization. Sect. 4 details the collisional contributions, including the computation of depolarization and polarization transfer rates in the multi-level case. The results and their physical interpretation are discussed in Sect. 5, while Sect. 6 summarizes the solar implications of our findings and presents the main conclusions. The relevant collisional data are provided in Appendix A.
 
\section{Generation of the atomic alignment in the O\,\textsc{i} levels}
Consider 
an atomic state of a complex atom like the O\,\textsc{i}  
denoted by $|\alpha \,  \mathcal{J} M_\mathcal{J}\rangle$, 
where $\mathcal{J}$ is the quantum number corresponding to total angular momentum ($\mathbfcal{J}$), $M_\mathcal{J}$ its projection  over  a quantization axis  aligned with the magnetic field, 
and 
$\alpha$ represents the set of additional quantum numbers required to fully specify the state.   
In 
the context of the polarization studies, 
the representation of  O\,\textsc{i}  states using the atomic density matrix formalism based on irreducible tensorial operators $\rho_q^{k_\mathcal{J}} (\alpha \,  \mathcal{J})$  has been  demonstrated to be the most appropriate  (e.g., Sahal-Br\'echot 1977; Landi Degl'Innocenti  \& Landolfi 2004). 
Here,
the index $k_\mathcal{J}$ denotes the tensorial order inside the level where
$0 \le k_\mathcal{J} \le 2 \mathcal{J}$
and $q$ quantifies the coherences between magnetic sublevels, with $-k_\mathcal{J} \le q \le +k_\mathcal{J}$.

We consider    an optically thin slab  in the solar atmosphere, composed of O\,\textsc{i} atoms,
illuminated 
anisotropically by an unpolarized photospheric radiation field. 
The incident radiation, 
with wavelength $\lambda = \frac{c}{\nu}$ 
where $c$ is the speed of light and $\nu$ its frequency, 
is 
assumed to exhibit cylindrical symmetry around the local solar vertical passing through the scattering point. 
Under this symmetry, 
only the multipole components with 
$k = 0$ (the mean intensity $J^0_0(\nu)$) 
and 
$k = 2$ (the radiation anisotropy component $J^2_0(\nu)$) 
are 
required to fully describe the incident radiation field.

In the case of an unpolarized and cylindrically symmetric incident radiation field, only atomic alignment can be induced, which corresponds to non-zero density matrix elements $\rho_q^{k_{\mathcal{J}}}$ with $q=0$ and even $k_{\mathcal{J}}$. This atomic alignment is the origin of the linear polarization that arises in the scattered radiation (e.g. Landi Degl’Innocenti  \& Landolfi 2004).

\begin{figure}[h]  
\begin{center}
\includegraphics[width=9 cm]{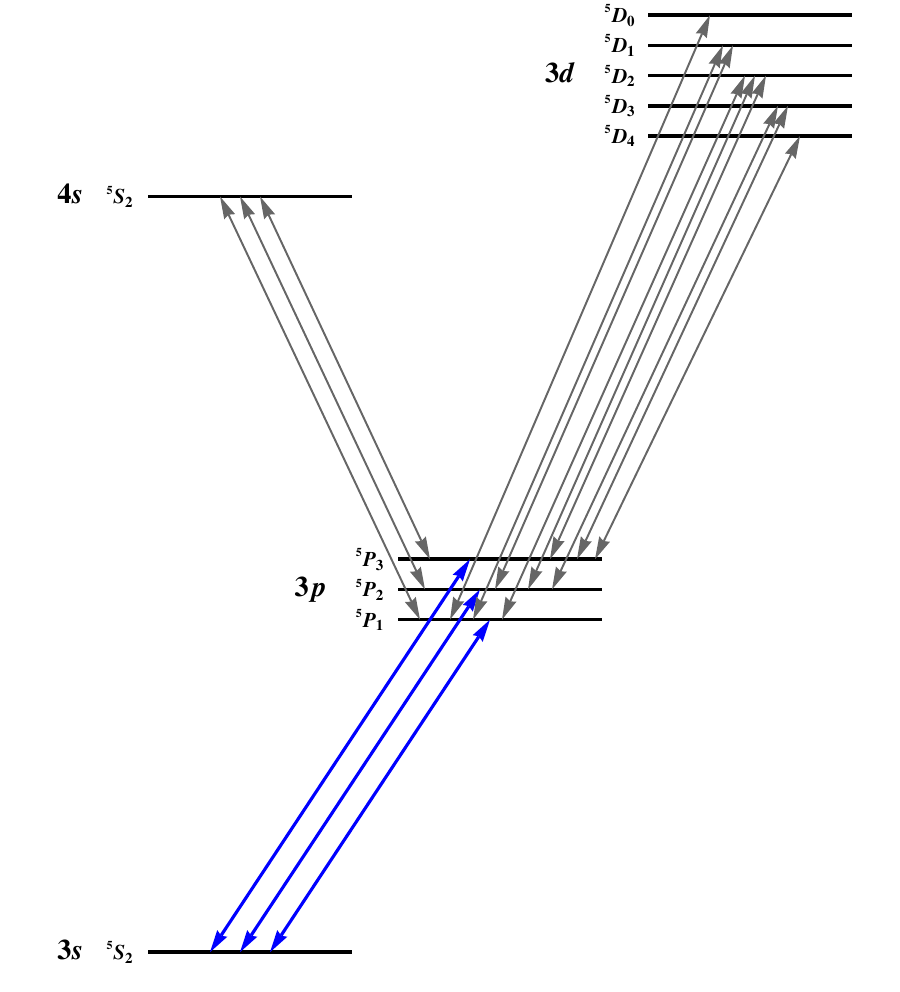}  
\caption{Energy level diagram of the atomic model of the O\,\textsc{i} IR triplet adopted in this work. 
Blue arrows indicate the transitions of the O\,\textsc{i} IR triplet, while grey arrows represent the other radiative transitions included in the model.}
\label{OI_Atomic_Model}
\end{center}
\end{figure}

The atomic model employed in this study 
(see Figure~\ref{OI_Atomic_Model}) consists of selected $s$-, $p$-, and $d$-states 
to describe the formation and polarization of the O\,\textsc{i} IR triplet around 777\,nm, corresponding to the $(3p\,^5P_{\mathcal{J}} \rightarrow 3s\,^5S_2)$ transitions   with $\mathcal{J}=1,2,3$ for the upper $3p\,^5P_{\mathcal{J}}$ levels.   
This model retains only the levels that are related to the generation of the 
O\,\textsc{i} IR triplet,
and 
leaves out higher-lying excited states due to their low population and/or weak coupling with the triplet levels under typical solar atmospheric conditions.
Some levels of comparable or lower energy are also omitted because their transitions to the included states are forbidden by the selection rules of electric–dipole transitions, making their radiative probabilities negligible. In addition, inelastic collisions with electrons are not expected to be strong enough to compensate for the radiative weakness of such transitions.

The scattering polarization of the O\,\textsc{i} IR triplet is the observational signature of atomic alignment, quantified by the $\rho^{K}_{Q}(J)$ multipole components of the density matrix (see, e.g., Landi Degl’Innocenti \& Landolfi 2004 for details on the physical link between alignment and polarization). Once these components are determined, the emergent fractional linear polarization $Q/I$ can be obtained from the standard relations that connect it to the degree of alignment (e.g., Trujillo Bueno 1999; Derouich et al. 2007). It is worth noting that these approximate relations were derived for \(90^\circ\) scattering in an optically thin slab. They are widely used for trend and sensitivity analyses  (e.g., Belluzzi et al.\ 2009; Belluzzi \& Trujillo Bueno 2011, including Ca II H\&K, Mg II h\&k, Na I D$_1$\&D$_2$, and O I 777\,nm). In the future, a more quantitative study will require full non-LTE radiative-transfer synthesis.

In addition to radiative rates associated with emission and absorption processes, we include two other main contributions in modeling the formation and polarization of the O\,\textsc{i} IR triplet lines:
isotropic collisions with neutral hydrogen atoms
and
the Hanle effect due to a turbulent magnetic field.
The radiative contribution to the SEE
adopted
in this study follows the formalism of Landi Degl’Innocenti \& Landolfi (2004), within the framework of the so-called multi-level atom in the tensorial basis.

In the numerical implementation, the contribution of the incident radiation field is specified through the anisotropy factor, defined as \(w = \sqrt{2}\, J^2_0/J^0_0\), and the number of photons per mode, given by \(\bar{n} = J^0_0 \left( c^2/2 h \nu^3 \right)\). These quantities are evaluated following the method described by Manso Sainz \& Landi Degl’Innocenti (2002) based on the data given by Cox (2000). The Einstein coefficients required for the radiative rates entering the SEE are extracted from the NIST database.

In the following sections, we describe the inclusion of the magnetic and collisional contributions in the  SEE.

\section{Magnetic contribution}
Adopting a frame whose quantization axis is aligned with the symmetry axis of the radiation field,  the Hanle effect of a magnetic-field  on the irreducible density–matrix components \(\rho^{k_{\mathcal J}}_{q}(\alpha\,\mathcal J)\) is given by  (e.g. Landi Degl’Innocenti et al.\ 1990):
\begin{equation} \label{eq_6}
\left(\!\! \frac{d\,\rho^{k_{\mathcal J}}_{q}(\alpha\,\mathcal J)}{dt} \!\!\right)_{\rm mag}
= -\,i\,\omega_{L}\,g_{\mathcal J}\sum_{q'} \mathcal K^{k_{\mathcal J}}_{q q'}\,
\rho^{k_{\mathcal J}}_{q'}(\alpha\,\mathcal J),
\end{equation}
where:
\begin{itemize}
  \item \(\omega_{L}\) is the Larmor frequency, proportional to the magnetic-field strength \(B\);
  \item \(g_{\mathcal J}\) is the Landé factor of level \((\alpha \mathcal J)\);
  \item \(\mathcal K^{k_{\mathcal J}}_{q q'}\) is the magnetic kernel that couples components with different \(q\)
        and accounts for the rotation from the magnetic frame (quantization axis along \(\mathbf B\)) to the chosen
        reference frame (see Landi Degl’Innocenti et al.\ 1990).
\end{itemize}
All radiative rates entering the SEE are formally invariant under rotations of the reference frame
(e.g. Landi Degl’Innocenti  \& Landolfi 2004).

When the magnetic field is turbulent, it is appropriate to average the effect of the magnetic field over all possible orientations. The angle-averaged density matrix component is:
\begin{equation} \label{eq:angle_avg_rho}
\overline{ \rho_q^{k_\mathcal{J}}}(\alpha \, \mathcal{J}) = \frac{1}{4\pi} \int_0^{2\pi} d\chi_B \int_0^\pi \sin \theta_B \, d\theta_B \, \rho_q^{k_\mathcal{J}} (\alpha \, \mathcal{J}; \theta_B, \chi_B),
\end{equation}

This integral accounts for all possible magnetic field directions $(\theta_B, \chi_B)$, assuming a constant field strength $B$, where $\theta_B$ is the inclination angle with respect to the quantization axis, and $\chi_B$ is the azimuthal angle measured in the plane perpendicular to that axis. 
With unpolarized, axially symmetric illumination and an isotropically distributed (turbulent) magnetic field, the \(q\neq 0\) coherences cannot be produced, leaving only the azimuthally invariant \(q=0\) components (independent of \(\chi_B\)).
By using the change of variables  $ \mu_{B}=\cos \theta_B$, Equation  \ref{eq:angle_avg_rho} becomes:
\begin{equation}
\overline{ \rho_0^{k_\mathcal{J}}} (\alpha \, \mathcal{J})=\frac{1}{2} \int_{-1}^{1} d\mu_B\, \rho_0^k(\alpha \, J; \mu_B) \, .
\end{equation}
$\rho_0^k(J; \mu_B)$ is the solution of the SEE for a given direction of the magnetic field, characterized by the inclination $\theta_B$ (since it is independent of $\chi_B$).
The averaging over $\theta_B \in [0, \pi]$ ensures isotropy of the field.
This treatment implicitly assumes that the magnetic field strength is constant but its direction is random.  The averaging over $\theta_B$ has been carried out using a Gaussian quadrature with $N=15$ points, which provides accurate convergence.

For the sake of simplicity, in Sections~5 and 6 we denote the isotropic-field-averaged density matrix element  
$\overline{\rho_0^{k_\mathcal{J}}}(\alpha \, \mathcal{J})$  simply as  $\rho_0^{k_\mathcal{J}}(\alpha \, \mathcal{J})$.

\section{Collisional contribution  in the multi-level case}
\subsection{SEE collisional contribution}
In the tensorial basis, the contribution of depolarizing isotropic collisions to the SEE is given by (e.g.  Sahal-Br\'echot et al. 2007): 
\begin{eqnarray} \label{eq_ch3_17}
\left(\!\! \frac{d \; \rho_q^{k_\mathcal{J}} (\alpha \, \mathcal{J})}{dt}   \!\!\right)_{\rm coll} 
\!\!&=&\!\! 
- D^{k_\mathcal{J}}(\alpha \, \mathcal{J}, T) \; \rho_q^{k_\mathcal{J}} (\alpha \, \mathcal{J}) \nonumber \\
&&\!\! \hspace{-2cm} -   \rho_q^{k_\mathcal{J}}  (\alpha \, \mathcal{\mathcal{J}}) 
\sum_{\mathcal{J}' \ne \mathcal{J}}  \sqrt{\frac{2\mathcal{J}'+1}{2\mathcal{J}+1}} D^0 (\alpha \, \mathcal{J} \to \alpha \, \mathcal{J}', T)    \\
&&\!\!  
\hspace{-2cm} + \sum_{\mathcal{J}' \ne \mathcal{J}}  
D^{k_\mathcal{J}}(\alpha \, \mathcal{J}' \to \alpha \, \mathcal{J}, T) \;  \rho_q^{k_\mathcal{J}} (\alpha \, \mathcal{J}') \nonumber \, ,
\end{eqnarray}
where 
$0 \!\le\! k_\mathcal{J} \!\le\! k_{\mathcal{J}, \rm max}$ 
with 
$k_{\mathcal{J}, \rm max}\!=\! 2 \mathcal{J}$ 
for $ \mathcal{J}\!=\! \mathcal{J}'$ 
and 
$k_{\mathcal{J}, \rm max}\!=\! min \{2 \mathcal{J},2 \mathcal{J}'\}$  
for  
$\mathcal{J} \!\ne\! \mathcal{J}'$.
Here $D^k(\alpha \, \mathcal{J}, T)$ 
and $D^k(\alpha \, \mathcal{J} \!\to\! \alpha \, \mathcal{J}', T)$  are  the depolarization and the polarization transfer rates, respectively.  
Note that 
$D^0(\alpha \, \mathcal{J}, T)$ = 0 
by definition and that 
$D^0 (\alpha \, \mathcal{J} \!\to\! \alpha \, \mathcal{J}', T)$ 
is the population transfer rate between the levels  $(\alpha \, \mathcal{J})$ and $(\alpha \, \mathcal{J}')$. 
These 
rates  should be  calculated independently.

\subsection{Collisional rates calculation}
To calculate the depolarization and transfer of polarization
rates 
for levels of complex atoms, Derouich et al. (2005a) 
proposed 
a collisional method based on the frozen core approximation (see also Derouich 2020). 
In the framework of this  method, 
the complex atom is
assumed to be composed of the following:
\begin{itemize}
\item Electrons in an internal subshell where their total orbital
momentum is zero.
\item Outer incomplete (open) shell containing electrons with
non-zero total orbital momentum denoted as $\bf L_c$ and a total
spin $\bf S_c$. This shell is called the core of the complex atom.
\item An optical electron with orbital angular momentum $\bf l$ and
spin $\bf s$ ($s=\frac{1}{2}$).  
\end{itemize}
We use the following coupling scheme to obtain the total angular momentum $\mathbfcal{J}$:
\begin{align*}
   {\bf j} &= {\bf l + s}, \\
   {\bf J_c} &= {\bf L_c + S_c}, \\
   {\mathbfcal{J}}   &= {\bf j + J_c},
\end{align*}
where:
\begin{itemize}
  \item $\bf l$ and $\bf s$ are the orbital and spin angular momenta of the valence electron ($l$ and $s$ are the corresponding quantum numbers),
  \item $\bf j$ is the total angular momentum of the optical electron ($j$ is the corresponding quantum number),
  \item $\bf L_c$ and $\bf S_c$ are the total orbital and spin angular momenta of the core ($L_c$ and $S_c$ are the corresponding quantum numbers),
  \item $\bf J_c$ is the total angular momentum of the core ($J_c$ is the corresponding quantum number),
  \item $\mathbfcal{J}$ is the total angular momentum of the complex atom which is here the O\,\textsc{i} ($\mathcal{J}$ is the corresponding quantum number).
\end{itemize}
We assume that only the valence electron is affected by the interaction with neutral hydrogen atoms. The  electrons of the core are considered to not be  influenced by the collisions, 
in the sense that $\bf L_c$, $\bf S_c$, 
and $\bf J_c = L_c + S_c$ are conserved under the frozen core approximation.
 Therefore, the polarization transfer rates
$D^{k_\mathcal{J}}(\alpha \, \mathcal{J}' \to \alpha \, \mathcal{J}, T) $ are given by the following
linear combination of the rates of simple atoms $D^{k_j}(j \rightarrow j') $ (see Nienhuis 1976; Omont 1977; Derouich  2020):
\begin{eqnarray}  \label{Eq_2}
D^{k_\mathcal{J}}(\alpha \, \mathcal{J} \rightarrow \alpha \, \mathcal{J}')  & = & (2 \mathcal{J}+1)  (2 \mathcal{J}'+1)   \\ 
&   & \hspace{-2cm}  \times \sum_{k_j}  (2k_{k_j}+1) D^{k_j}(j \rightarrow j') \nonumber \\ 
&   & \hspace{-2cm} 
\times
\sum_{k_{J_c}} (2k_{J_c}+1) \left\{ \begin{array}{ccc}
j & J_c & \mathcal{J} \\
j & J_c & \mathcal{J} \\
k_j & k_{J_c} & k_\mathcal{J} \end{array} \right\}  \left\{ \begin{array}{ccc}
j' & J_c & \mathcal{J}' \\
j' & J_c & \mathcal{J}' \\
k_j & k_{J_c} & k_\mathcal{J} \end{array} \right\} \nonumber \, ,
\end{eqnarray}
and, for the depolarization rates, one has:
\begin{eqnarray}   \label{Eq_3}
D^{k_\mathcal{J}}(\alpha \, \mathcal{J})  & = & (2 \mathcal{J}+1)^2     \sum_{k_j}  (2k_{k_j}+1) D^{k_j}( j)   \nonumber \\ 
&   &  \times
\sum_{k_{J_c}} (2k_{J_c}+1) \left\{ \begin{array}{ccc}
j & J_c & \mathcal{J} \\
j  & J_c & \mathcal{J} \\
k_j  & k_{J_c} & k_\mathcal{J} \end{array} \right\}^2 \, .
\end{eqnarray}

The atomic model  of the O\,\textsc{i}  adopted in this work involves levels where the orbital momentum of the valence electron   is $l=0, 1,$ or $2$. 
Rates $D^{k_j}(j \rightarrow j')$ and $D^{k_j}(j)$  
for the cases where $l=1$ or $2$   (i.e., $p-$  and $d-$states)
are 
tabulated in the form of variation laws for all orders $k_j$ in Derouich (2020). 
For $l=0$, $D^{k_j}(j)$ 
are also available in the form of variation laws for all s-states in Derouich et al (2005b). 
These 
variation laws depend on the effective quantum number $n^{*}$,
which must be calculated 
for all levels involved in the atomic model.
Once $n^{*}$ is determined,
the corresponding  $D^{k_j}(j \rightarrow j')$ and $D^{k_j}(j)$ rates can be obtained, 
and subsequently used to
calculate the rates 
$D^{k_\mathcal{J}}(\alpha  \, \mathcal{J} \rightarrow \alpha  \, \mathcal{J}') $ 
and 
$D^{k_\mathcal{J}}(\alpha \, \mathcal{J})$  of the O\,\textsc{i} atoms 
due to collisions with hydrogen atoms.  
Recall that, 
if the energy of the level of interest is
denoted by $E_{\rm level}$, and the ionization energy of the perturbed
atom in its fundamental state is 
$E_{\infty}$, then the effective quantum number is given by 
$$n^{*}=[2(E_{\infty}-E_{\rm level})]^{-1/2} \, ,$$
where both 
$E_{\infty}$ 
and 
$E_{\rm level}$ 
are expressed in atomic units. 
\begin{figure}[h]  
\begin{center}
\includegraphics[width=8.5 cm]{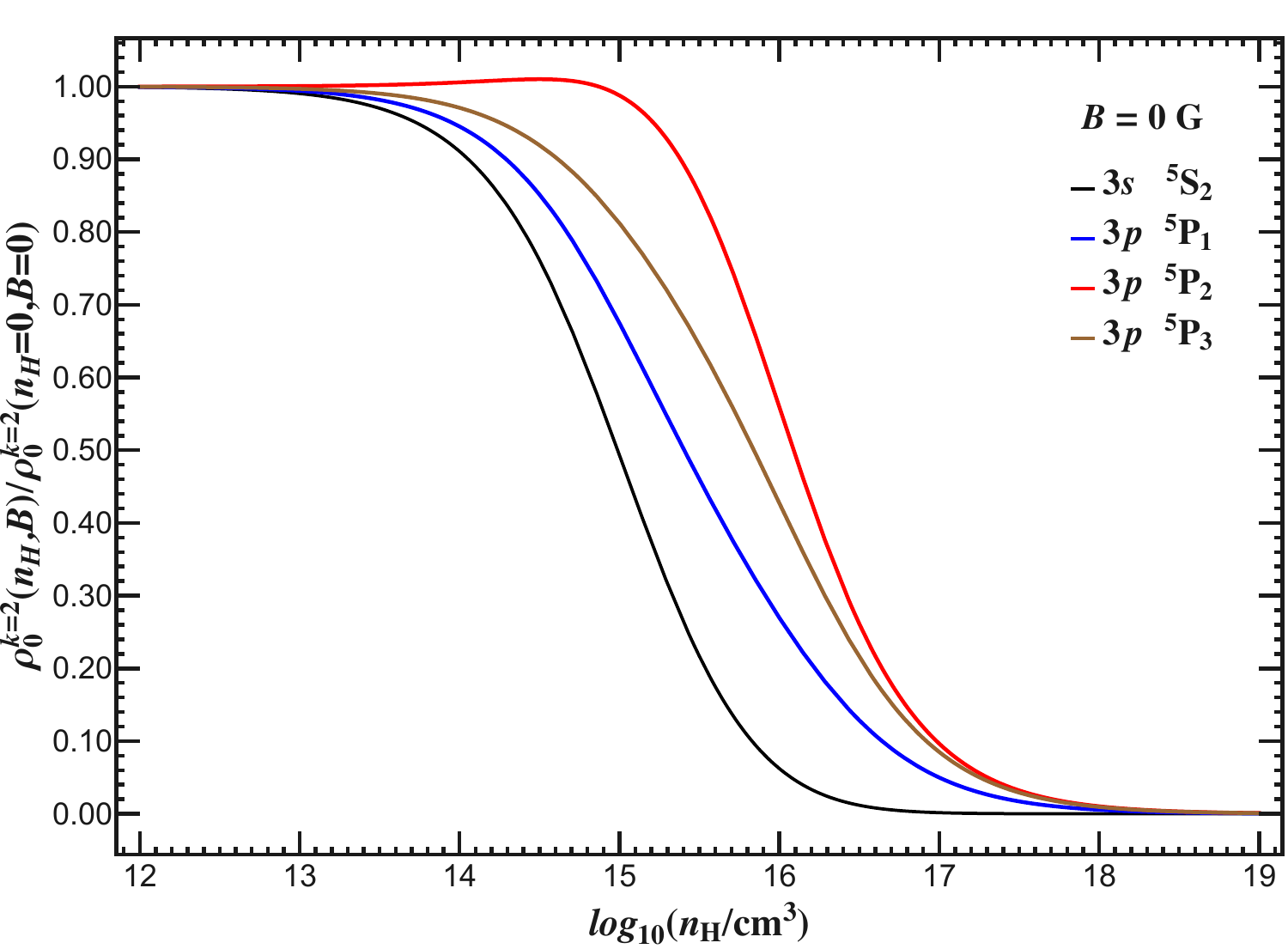}  
\caption{Normalized atomic alignment  $[{\rho_0^{(2)}(n_{\mathrm{H}})}/{\rho_0^{(2)}(n_{\mathrm{H}} \!=\! 0)}]_{B = 0}$ as a function of hydrogen density $n_{\mathrm{H}}$ in the absence of a magnetic field ($B \!=\! 0$), showing the pure effect of elastic collisions on the fine-structure levels of the O\,\textsc{i} IR triplet.}
\label{OI_collisions_sensitivity_B0}
\end{center}
\end{figure}
Following the procedure outlined above, we calculate the rates $D^{k_\mathcal{J}}(\alpha \, \mathcal{J})$ and $D^{k_\mathcal{J}}(\alpha \, \mathcal{J} \rightarrow \alpha \, \mathcal{J}')$, and express them in the following form:
\begin{eqnarray} \label{eq_MT}
D^{k_\mathcal{J}}=a^{k_\mathcal{J}} \times 10^{-9}  \; n_H   \left(\! \frac{T}{5000} \!\right)^{\lambda^{k_\mathcal{J}}} \, .
\end{eqnarray} 
The coefficients $a^{k_\mathcal{J}}$, corresponding to depolarization and polarization transfer rates, are listed in Tables~\ref{table_1} and~\ref{table_2} of Appendix A, respectively, for all levels of the model of O\,\textsc{i} IR triplet under consideration and for all even orders $k_\mathcal{J}$\footnote{Recall that odd orders of $k$ are not relevant in the present context, as only linear polarization is considered in this study.}. 
As explained in Derouich et al. (2005b)  and  in Derouich  (2020), 
$\lambda^{k_\mathcal{J}}$ can be approximated by $0.416$ for $s$-states and by $0.38$ for $p$- and $d$-states. 
 We specify that the adopted $D^{k_\mathcal{J}}$ rates are calculated in the absence of magnetic fields (i.e., zero-field rates), which should be a good approximation under the Hanle-effect regime field strengths   considered here, since the impact of the field on the interaction potential is expected to be weak and  $\mathcal{J}$ remains a good quantum number (see Derouich \& Qutub 2024 for details).   

It is worth noting that  radiative rates are taken from available databases where they are usually computed within the L-S coupling scheme, wheras collisional rates are calculated using a different angular momentum coupling. Although the two coupling schemes are not formally identical, this mixed approach is common practice and allows us to make use of the best available data for both radiative and collisional processes.

\section{Results and discussion}
\subsection{Effect of  collisions}
Using  the multi-level atomic model for O\,\textsc{i} IR triplet shown in Figure~\ref{OI_Atomic_Model},  we  solve the SEE to compute the non-zero components $ \rho_0^{k_\mathcal{J}} $ in the presence of a turbulent magnetic field.
The numerical code incorporates the collisional rates given by Equation~\ref{eq_MT} with the values of the $a^{k_\mathcal{J}}$ coefficient provided in Tables~\ref{table_1} and~\ref{table_2} as inputs, 
along with an anisotropic radiation field characterized by $J^0_0(\nu)$ and $J^2_0(\nu)$.  
These 
computations are performed over a wide range of hydrogen density, from  $10^{11}$ cm$^{-3}$ to  $10^{19}$ cm$^{-3}$, assuming a solar temperature of $T= 5000$ K, and magnetic field strengths in the range $1$ mG to $100$ G.

Since collisional rates are proportional to the hydrogen density $n_H$,  
as per the impact approximation, 
studying the effect of collisions on polarization is effectively equivalent to analyzing its dependence on $n_H$. 
Figure~\ref{OI_collisions_sensitivity_B0} 
illustrates the normalized atomic alignment 
$[{\rho_0^{(2)}(n_{\mathrm{H}})}/{\rho_0^{(2)}(n_{\mathrm{H}} \!=\! 0)}]_{B=0}$ 
as a function of the hydrogen density $n_{\mathrm{H}}$ 
for each of the four fine-structure levels involved in the O\,\textsc{i} IR triplet: 
the upper levels $3p\,^5P_1$ (blue curve), $3p\,^5P_2$ (red curve), $3p\,^5P_3$ (brown curve), and the lower level $3s\,^5S_2$ (black curve).  
The figure shows that all levels begin to be sensitive  
to elastic collisions with neutral hydrogen for  $n_{\mathrm{H}} \!\gtrsim\! 10^{14}$ cm$^{-3}$.  
Among them, the $3s\,^5S_2$ level 
depolarizes more rapidly than the upper levels,
becoming
almost completely depolarized around 
$n_{\mathrm{H}} \!\sim\! 10^{16}\,\mathrm{cm}^{-3}$, 
which is typical of photospheric conditions. In contrast, at chromospheric densities 
($n_{\mathrm{H}} \!\sim\! 10^{14}\,\mathrm{cm}^{-3}$), 
its alignment is only slightly affected by collisions.

Focusing on the upper levels, the state $3p\,^5P_2$ retains its alignment more effectively than the states
$3p\,^5P_1$ and $3p\,^5P_3$, 
maintaining approximately 70\% of its alignment even at $n_{\mathrm{H}} \!\sim\! 10^{16}\,\mathrm{cm}^{-3}$.
By comparison, the alignment of $3p\,^5P_1$ (the most sensitive to depolarizing collisions among the upper levels)
and 
$3p\,^5P_3$ decreases more rapidly with increasing 
$n_{\mathrm{H}}$, 
though remains noticeable even at $n_{\mathrm{H}} \!\sim\! 10^{16}\,\mathrm{cm}^{-3}$. 
Thus,
in the absence of magnetic fields, 
the upper levels of O\,\textsc{i} IR lines
(especially $3p\,^5P_2$)
remain polarized at photospheric densities ($n_{\mathrm{H}} \!\sim\! 10^{15}$--$10^{16}\,\mathrm{cm}^{-3}$), as shown in Figure~\ref{OI_collisions_sensitivity_B0}.

\begin{figure}[h!]  
\begin{center}
\includegraphics[width=8.75 cm]{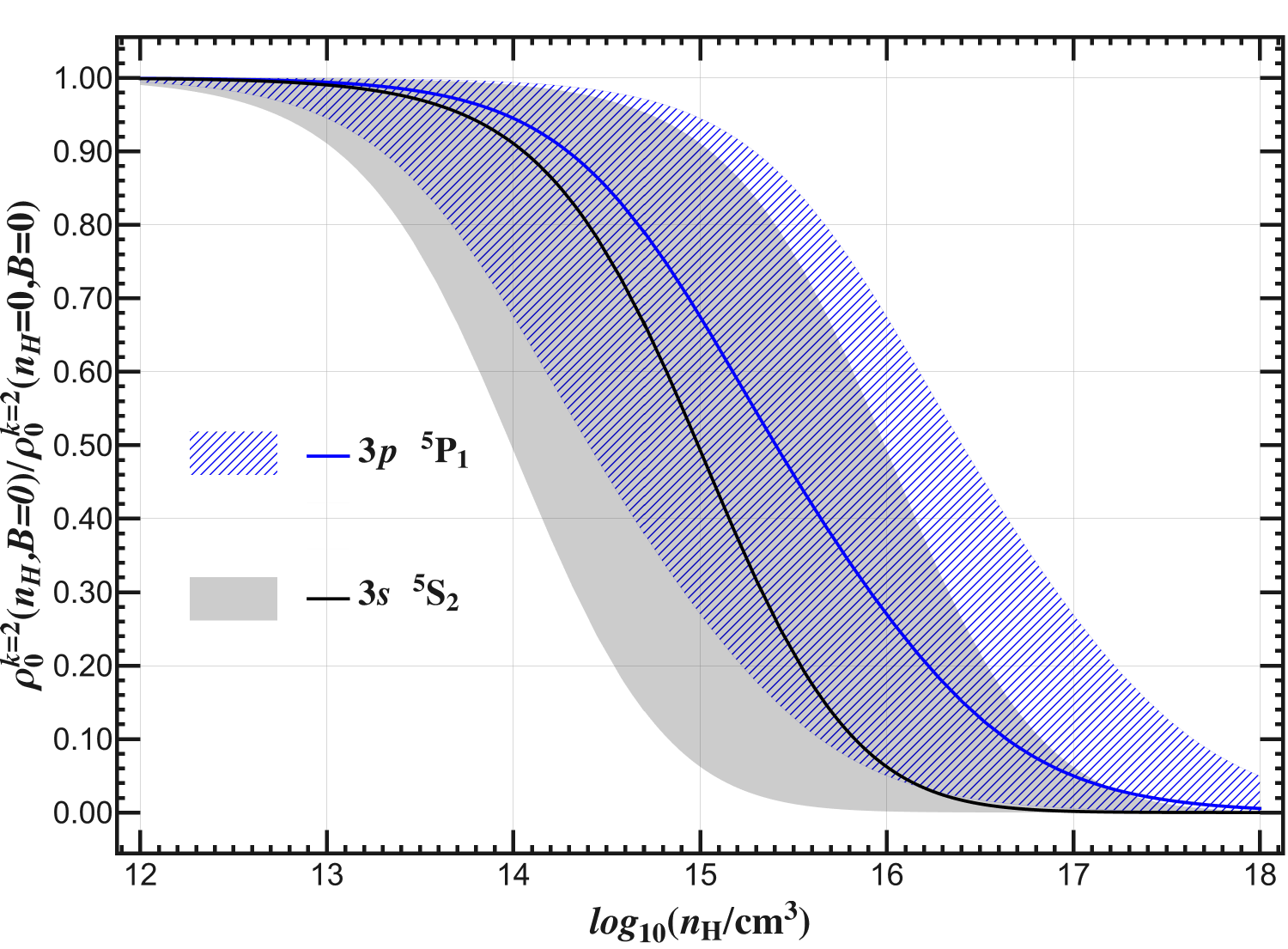}  
\caption{Sensitivity of the normalized alignment,
\(\rho^{2}_{0}(n_{\mathrm H},B=0)/\rho^{2}_{0}(n_{\mathrm H}=0,B=0)\),
to uncertainties in the elastic O--H collisional operator \(D\) as a
function of \(\log_{10}(n_{\mathrm H}/\mathrm{cm^{3}})\).
For each level, the solid curve is the reference calculation (\(D\)),
while the shaded band spans the range obtained by uniformly scaling all
elastic rates to \(D/10\) and \(10  \times D\). The gray band and black
line refer to the \(S_{2}\) level (3s \(^{5}\!S\), \(J=2\)), and the blue
band and line to the \(P_{1}\) level (3p \(^{3}\!P\), \(J=1\)) shown in
Fig.~2. Results are for \(B=0\). }
\label{percentage_sensitivity}
\end{center}
\end{figure}

To explore the impact of large uncertainties, which may arise from the use of rough approximations in calculating the elastic collisional rates for collisions between hydrogen and oxygen atoms, we have tested the effect of increasing or decreasing the rates by one order of magnitude. Here, $D$ denotes the total depolarizing effect of elastic H--O collisions, represented by the complete set of depolarization and polarization transfer rates listed in Tables~A.1 and A.2. We express the results as the percentage variation of the alignment relative to the reference case (1 $\times$ $D$), obtained by either decreasing the rates by a factor of 10 ($D/10$) or increasing them by a factor of 10 (10 $\times$ $D$). These percentage variations can take both negative and positive values, depending on whether the alignment decreases or increases relative to the reference case. At $n_{\mathrm{H}} = 10^{15}\,\text{cm}^{-3}$, the alignment of the 3s $^{5}S_2$  level changes from --84.7\% ($D/10$) to +87.4\% (10 $\times$ $D$), while the alignment of the 3$p$ $^{3}P_1$ level varies from --40.0\% to +60.1\%. At $10^{16}\,\text{cm}^{-3}$, the discrepancies become even larger, reaching --692.7\% and +97.4\% for the 3$s$ $^{5}S_2$ level, and --150.4\% and +81.5\% for the 3$p$ $^{3}P_1$ level. At $10^{17}\,\text{cm}^{-3}$,   where the alignment of 3s $^{5}S_2$  is already very small,  the deviations are extreme, with the 3s $^{5}S_2$ level ranging from --3816.6\% to +98.8\% and the 3$p$ $^{3}P_1$ level from --440.2\% to +88.9\%.  Figure \ref{percentage_sensitivity}   illustrates the results.  We also performed the same analysis for the other two upper levels of the triplet, 3$p$ $^{3}P_2$  and 3$p$ $^{3}P_3$, and found analogous results.   
This suggests that the O\,\textsc{i} IR  triplet polarization is highly sensitive to the adopted collisional rates.

\subsection{Interplay between collisions and magnetic fields}

\begin{figure}[h!]  
\begin{center}
\includegraphics[width=6.75 cm]{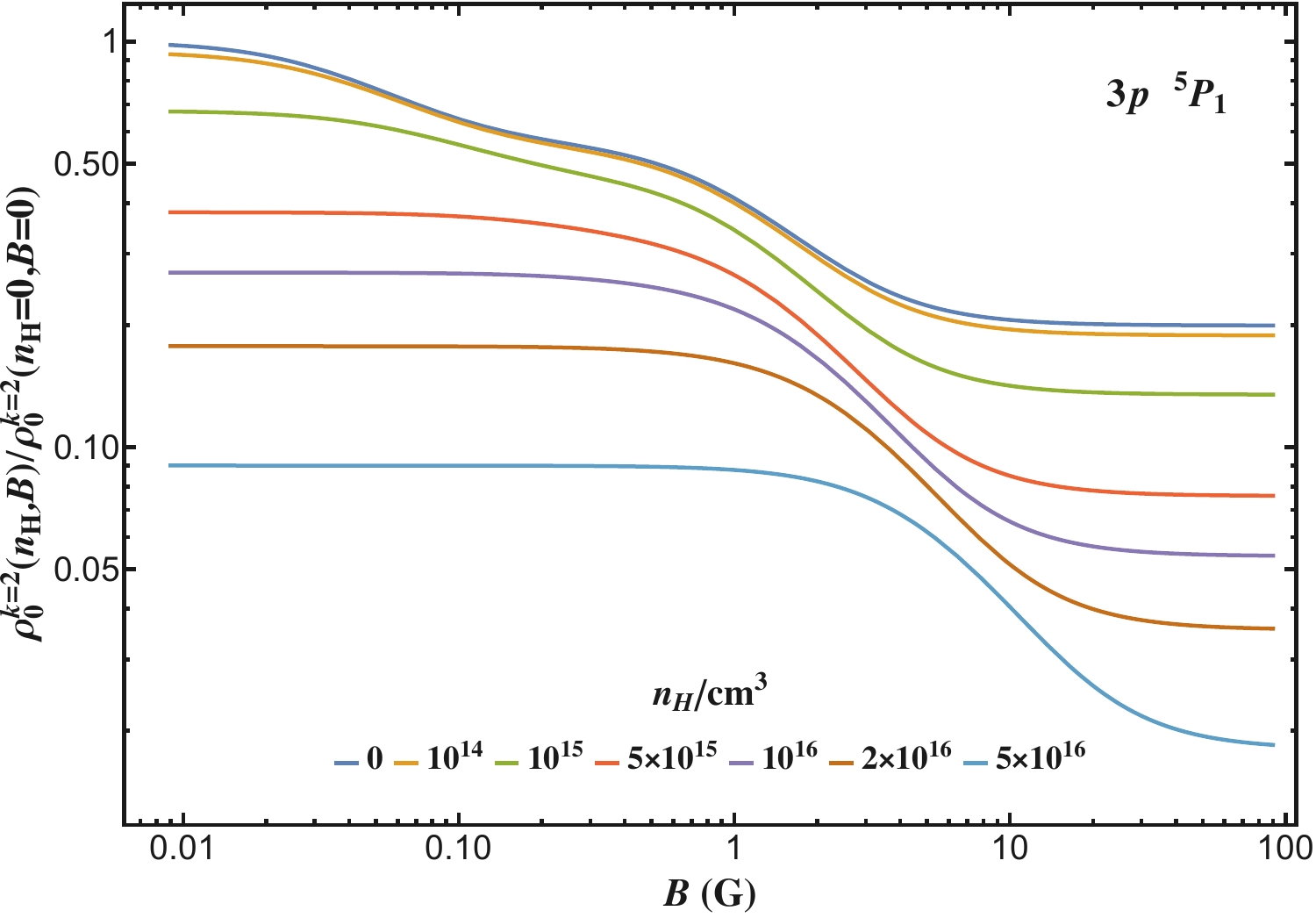}  
\includegraphics[width=6.75 cm]{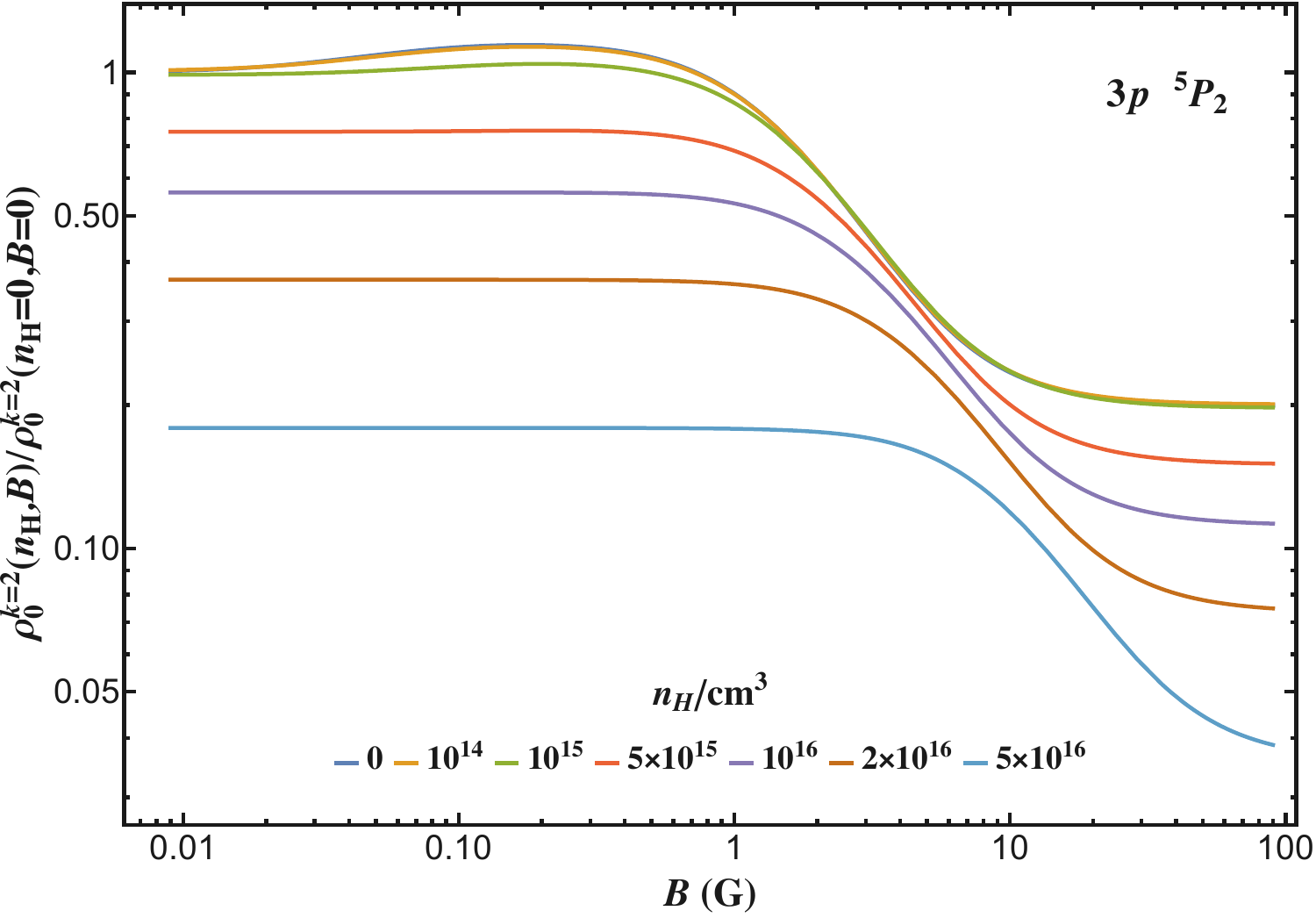}  
\includegraphics[width=6.75 cm]{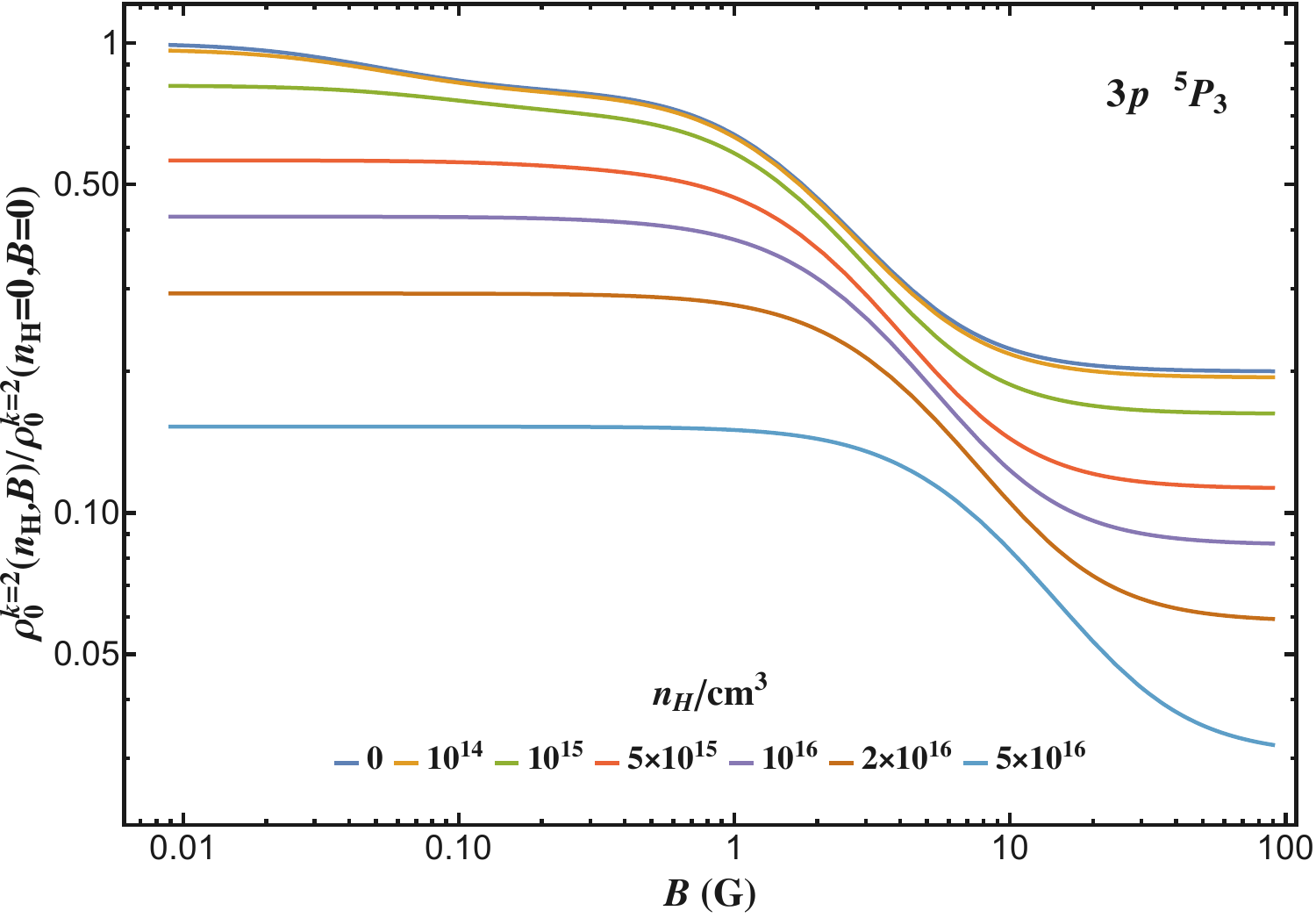} 
\includegraphics[width=6.75 cm]{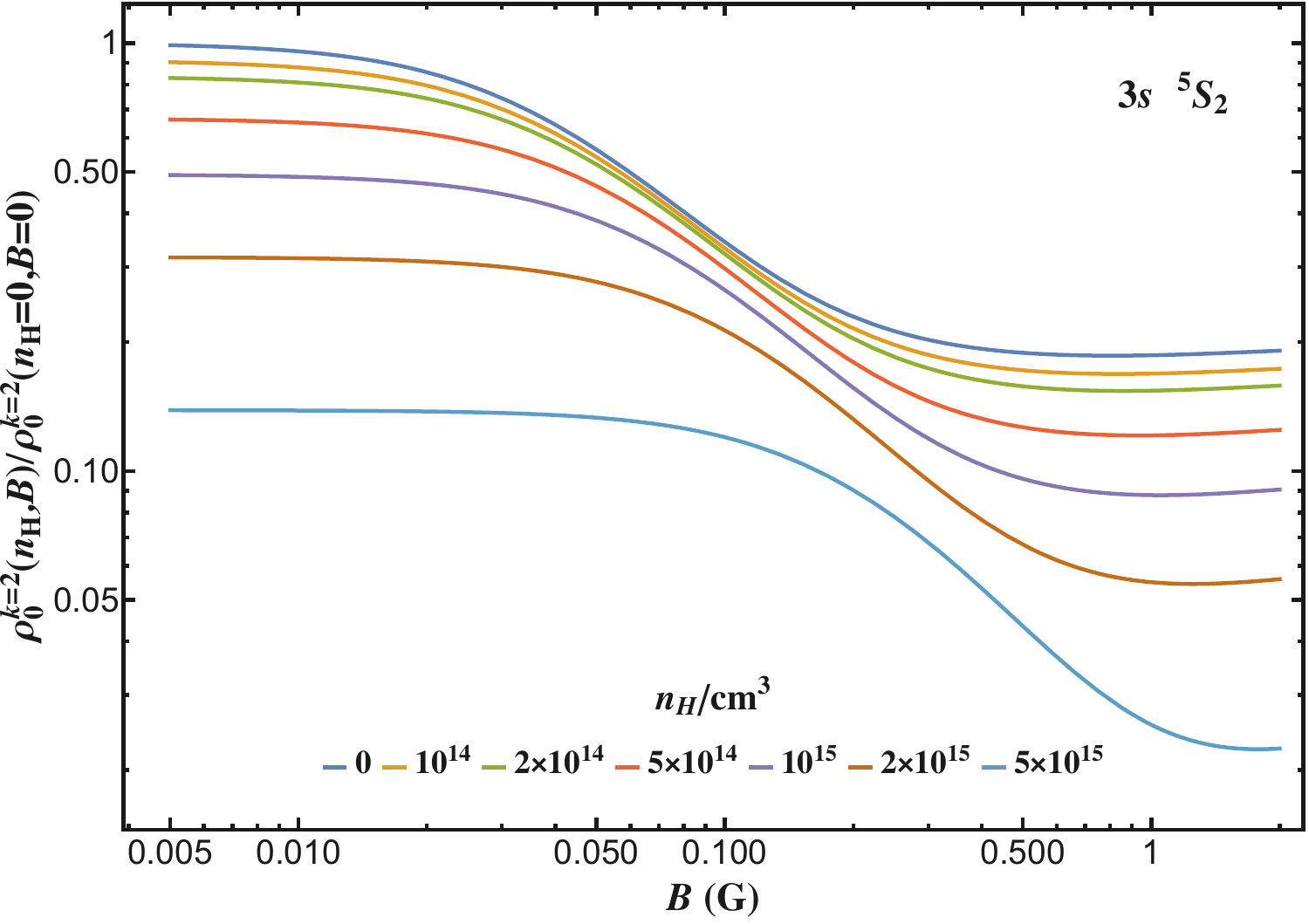} 
\caption{Normalized alignment, $\rho^{(2)}_0(n_{\mathrm{H}},B)/\rho^{(2)}_0(n_{\mathrm{H}}\!=\!0, B\!=\!0)$, as a function of the magnetic field strength $B$ for several hydrogen densities $n_{\mathrm{H}}$, for four levels of O\,\textsc{i}. From top to bottom: $3p\,^5P_1$, $3p\,^5P_2$, $3p\,^5P_3$, and $3s\,^5S_2$. Each curve corresponds to a fixed value of $n_{\mathrm{H}}$, ranging from $0$ to $5 \times 10^{16}\,\mathrm{cm}^{-3}$, as labeled. The plots illustrate how both elastic collisions with neutral hydrogen and the Hanle effect contribute to the depolarization of each level. }
\label{alignment_sensitivity}
\end{center}
\end{figure}

To assess the joint influence of the turbulent magnetic field and collisions on the atomic alignment of the O\,\textsc{i}  levels, 
we compute the normalized atomic alignment 
$\rho^{(2)}_0(n_\mathrm{H},B) / \rho^{(2)}_0(n_\mathrm{H}, B\!=\!0)$ as a function of the magnetic field strength $B$, for various fixed hydrogen densities $n_\mathrm{H}$. 
The results for $3p\,^5P_1$, $3p\,^5P_2$, $3p\,^5P_3$, and $3s\,^5S_2$ levels, presented in Figure~\ref{alignment_sensitivity}, show that the magnetic field plays a significant role in depolarizing the levels involved in the formation of the IR triplet. This highlights the interplay between the Hanle effect and elastic collisions in shaping the polarization signals.

\begin{figure}[h!]  
\begin{center}
\includegraphics[width=8.5 cm]{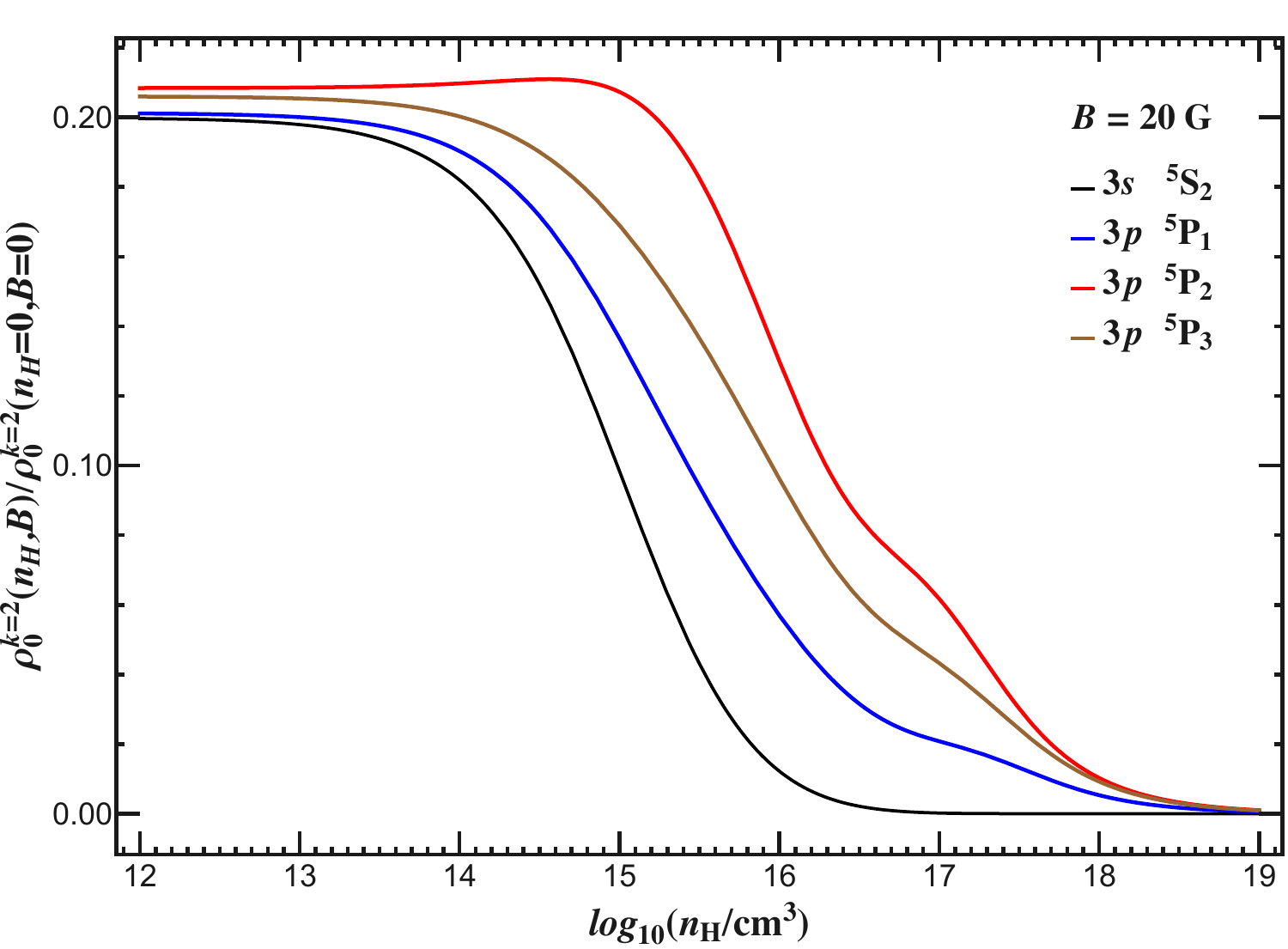}  
\caption{Same as Figure \ref{OI_collisions_sensitivity_B0}, but for a magnetic field strength of $B = 20$\,G, illustrating the combined effect of elastic collisions and magnetic depolarization (Hanle effect) on the atomic alignment of the O\,\textsc{i} IR triplet levels.}
\label{OI_collisions_sensitivity}
\end{center}
\end{figure}

The top three panels of Figure~\ref{alignment_sensitivity}  correspond to the upper levels of the IR triplet: 
$3p\,^5P_1$, $3p\,^5P_2$, and $3p\,^5P_3$, while the bottom panel corresponds to the lower level $3s\,^5S_2$. 
Each curve in the plots represents a different hydrogen density, ranging from $n_{\mathrm{H}} = 0$ to $5 \!\times\! 10^{16}\,\mathrm{cm}^{-3}$. Across all levels, the alignment decreases almost monotonically with increasing $B$ and $n_{\mathrm{H}}$, reflecting the progressive depolarization induced by elastic collisions and the Hanle effect.

The $3s\,^5S_2$ level is the most sensitive to depolarizing mechanisms:   for $n_{\mathrm{H}} \!\sim\!  10^{16}\,\mathrm{cm}^{-3}$ and $B \!\sim\!  20$\,G,   its alignment is almost entirely suppressed.
Among the upper levels, 
$3p$ $^5P_1$ is the most sensitive to magnetic depolarization, 
with a rapid decline in alignment for 
$B \!\gtrsim\! 1$~G 
when 
$n_\mathrm{H} \!\gtrsim\! 10^{15}$~cm$^{-3}$. 
In contrast, the alignment of  level $3p$  $^5P_3$ declines a bit more gradually, 
indicating moderately lower sensitivity to weak fields, 
and it is also less affected by collisional depolarization compared to 
$3p$  $^5P_1$.
 The alignment of level $3p\,^5P_2$ is the most resistant   among the upper levels, 
retaining a significant fraction of its alignment at $n_{\mathrm{H}} \!\lesssim\! 5 \!\times\! 10^{15}\,\mathrm{cm}^{-3}$ and $B \!\lesssim\! 10$\,G. 
It is also noteworthy that the magnetic field strength at which polarization is most sensitive shifts to higher values with increasing hydrogen density $n_\mathrm{H}$, indicating a coupled dependence on both collisional and magnetic depolarization mechanisms.

Complementing this analysis, Figure~\ref{OI_collisions_sensitivity} shows the normalized atomic alignment, 
defined as ${\rho^2_0(n_{\mathrm{H}}, B)}/{\rho^2_0(n_{\mathrm{H}} \!=\! 0, B \!=\! 0)}$,
as a function of hydrogen density for the same four O\,\textsc{i} levels for a magnetic field strength of $B = 20$\,G. This quantity captures how collisional and magnetic depolarization, acting together, reduce the alignment relative to the undisturbed case.
The lower level $3s\,^5S_2$ (black curve) 
depolarizes the fastest and most significantly: 
at the photospheric density $n_{\mathrm{H}} \!\sim\! 10^{15}\,\mathrm{cm}^{-3}$ 
and for a magnetic field strength of $B \!=\! 20$\,G, 
its alignment is practically zero. 
The alignment of the upper levels, 
$3p\,^5P_1$ (blue curve), $3p\,^5P_2$ (red curve), and $3p\,^5P_3$ (brown curve), is reduced by approximately 90\% at $n_{\mathrm{H}} \!\sim\! 10^{16}\,\mathrm{cm}^{-3}$ and $B \!=\! 20$\,G.  Stronger magnetic fields ($B \!>\! 20$\,G) cause alignment to decay even more rapidly (see Figure~\ref{alignment_sensitivity}), highlighting the combined depolarizing effects of the Hanle effect and collisions. 

Our analysis shows that incorporating collisional rates computed for photospheric hydrogen densities and a quiet-Sun turbulent magnetic field near the Hanle saturation regime into the SEE strongly suppresses atomic alignment.  This contrasts with the conclusion of del Pino Alemán \& Trujillo Bueno (2015, 2017), who reported that the bulk of the polarization is   destroyed in the photosphere, even under the action of very weak magnetic fields, well before the saturation regime is reached. A possible reason for this discrepancy is that their modeling relied on overestimated inelastic collisional rates that contribute to depolarization (whereas in our case they are negligible). In practice, this would means that part of the depolarization attributed in our work to the Hanle effect was, in their case, effectively taken up by inelastic collisions. Overestimating the impact of inelastic collisions, thus, would reduce   the magnetic field strengths required in their modeling, and this may have been compounded by their use of an approximate treatment of elastic collisions.

The 
rapid loss of alignment at $n_{\mathrm{H}} \!\gtrsim\! 10^{15}\,\mathrm{cm}^{-3}$,  
under magnetic field strengths typical of  the quiet-Sun, 
confirms that  a significant contribution to the  polarization of the IR triplet should primarily  originate in higher chromospheric layers (see also del Pino Alem\'an \& Trujillo Bueno  2015, 2017), where hydrogen densities are lower and collisional depolarization is weaker. In the future, it will be of great interest to incorporate accurate collisional depolarization rates into realistic non-LTE radiative transfer  calculations.

\section{Solar implications and conclusions}

Our study provides insights into the expected roles of elastic and inelastic collisions and turbulent magnetic fields in the formation and observability of the scattering polarization signals of the O\,\textsc{i} IR triplet in the solar atmosphere.

In the solar photosphere, where hydrogen densities are typically $n_\mathrm{H} \!\sim\! 10^{15}-10^{16}\,\mathrm{cm}^{-3}$, collisions alone lead to a substantial reduction of alignment, particularly for the lower level $3s\,^5S_2$ and the upper level $3p\,^5P_1$.   Among the upper levels, $^5P_2$ is found to retain the highest alignment.  The inclusion of a turbulent magnetic field with strength $B \sim 20$\,G further enhances this depolarization, leading to suppression levels above 90\% for most levels.
Conversely, in the upper chromosphere, where \(n_{\mathrm H}\lesssim 10^{14}\,\mathrm{cm^{-3}}\),
collisional depolarization becomes inefficient; even in the presence of magnetic fields,
a residual atomic alignment can persist and may be sufficient to produce measurable \(Q/I\) signals (see also del Pino Aleman \& Trujillo Bueno, 2015, 2017).

These results highlight the importance of  an accurate treatment of both collisional and magnetic depolarization mechanisms in non-LTE radiative transfer models to ensure accurate magnetic diagnostics based on the Hanle effect. The O\,\textsc{i} IR triplet remains a valuable probe of chromospheric magnetism, provided that the relevant physical processes are fully taken into account.

\section*{Acknowledgements}
This research work was funded by Institutional Fund
Projects under grant no. (IFPIP:52-130-1443). The authors gratefully acknowledge technical and financial support provided by the Ministry of Education and
King Abdulaziz University, DSR, Jeddah, Saudi Arabia.


\newpage
\begin{appendix}
\section{Collisional data} \label{coll_data}
\begin{table}[h!]
\begin{center}
\begin{tabular}{l c c c c c c r}
\hline
\hline
Term  &$\mathcal{J}$ &  $k$&  $a^k$    \\
 \hline 
2$s^2$ \; 2$p^3$ ($^4S$) $(3s)$  $^5S$     &  2         &  2  &  0.966         	 \\
    &  2         &  4  &  1.718         	 \\
2$s^2$ \; 2$p^3$ ($^4S$) $(4s)$  $^5S$    &  2       &  2  &   1.588    	 \\
      &  2         &  4  &  2.824         	 \\
  2$s^2$ \; 2$p^3$ ($^4S$) $(3p)$  $^5P$     &  1         &  2  &   1.781        	 \\
      &  2         &  2  &  1.723         	 \\
      &  2         &  4  &  1.903         	 \\
      &  3         &  2  &       1.355     	 \\
    &  3         &  4  &        2.094    	 \\
     &  3         &  6  &           1.744	 \\
 2$s^2$ \; 2$p^3$ ($^4S$) $(3d)$  $^5D$      &  1         &  2  &    1.361      	 \\
       &  2         &  2  &    1.227     	 \\
     &  2         &  4  &    1.366      	 \\
       &  3         &  2  &     1.249      	 \\
      &  3         &  4  &     1.293      	 \\
     &  3         &  6  &    1.369       	 \\
     &  4         &  2  &  1.083 \\
     &  4         &  4  &   1.395       	 \\
     &  4         &  6  &      1.316    	 \\
    &  4        &  8  &       1.398   	 \\
  \hline  
\end{tabular}
\end{center}
\caption{\textsf{O\,\textsc{i}   non-zero collisional depolarization rates $D^{k_\mathcal{J}} (\alpha \, \mathcal{J})$  for the multi-level case.}}
\label{table_1}
\end{table} 
\begin{table}[h!]
\begin{center}
\begin{tabular}{l c c c c c c r}
\hline
\hline
Term &$\mathcal{J}$ & $\mathcal{J}'$ &  $k$&  $a^k$    \\
 \hline 
  2$s^2$ \; 2$p^3$ ($^4S$) $(3p)$  $^5P$     &  1   &   2   &  0  &   0.744     \\
                                                                       &  1   &   2   &  2  &     0.113  	 \\
                                                                       &  1   &   3  &  0  &     0.655      \\
                                                                       &  1   &   3  &  2  &     0.321      \\
                                                                       &  2   &   3  &  0  &     1.196      \\
                                                                       &  2   &   3  &  2  &    0.780       \\
                                                                       &  2   &   3  &  4  &    0.146       \\
  2$s^2$ \; 2$p^3$ ($^4S$) $(3d)$  $^5D$     &  4   &   3  &  0  &      0.730     \\                 
                                                                       &  4   &   3  &  2  &        0.895   \\      
                                                                       &  4   &   3  &  4  &      1.066     \\     
                                                                       &  4   &   3  &  6  &     -0.026      \\          
                                                                       &  4   &   2  &  0  &       -0.197    \\                 
                                                                       &  4   &   2  &  2  &     -0.333      \\      
                                                                       &  4   &   2  &  4  &    0.757       \\     
                                                                       &  4   &   1  &  0  &      0.535     \\                 
                                                                       &  4   &   1  &  2  &      -0.737     \\      
                                                                        &  4   &   0  &  0  &      0.705     \\            
                                                                        &  3   &  2  &  0  &    1.632       \\            
                                                                        &  3   &  2  &  2  &      0.411     \\      
                                                                         &  3   &  2  &  4  &       0.265    \\    
                                                                          &  3   &  1  &  0  &   -0.014        \\      
                                                                         &  3   &  1  &  2  &     0.297      \\    
                                                                         &  3   &  0  &  0  &     -0.776      \\  
                                                                        &  2   &  1  &  0  &      0.159     \\ 
                                                                        &  2   &  1  &  2  &      0.367     \\ 
                                                                        &  2   &  0  &  0  &    0.006       \\      
                                                                        &  1   &  0  &  0  &    0.858       \\                                                  
                                            
  \hline  
\end{tabular}
\end{center}
\caption{\textsf{O\,\textsc{i}   collisional depolarization rates $D^{k_\mathcal{J}} (\alpha \, \mathcal{J} \to \alpha \, \mathcal{J'})$ for the multi-level case.}}
\label{table_2}
\end{table} 
\end{appendix}


\begin{thebibliography}{}

\bibitem[Asplund et al.(2021)]{Asplund2021} Asplund, M., Amarsi, A.M., \& Grevesse, N., 2021, A\&A, 653, A141.

\bibitem[Barklem(2007)]{Barklem2007} 
 Barklem, P. S. 2007, A\&A, 462, 781 
 
\bibitem[Belluzzi(2009)]{Belluzzi2009}
Belluzzi, L., Landi Degl’Innocenti, E., \&  Trujillo Bueno, J. 2009, APJ, 705:218--225.


\bibitem[Belluzzi(2011)]{Belluzzi2011}
Belluzzi, L., \&  Trujillo Bueno, J. 2011, APJ, 743:3 22 pp.
 

\bibitem[Cox (2000)]{Cox2000} Cox, A. N. 2000,  Allen's astrophysical quantities, 4th ed. Publisher: New York: AIP Press; Springer, 2000. Editedy by Arthur N. Cox. ISBN: 0387987460 

\bibitem[Del Pino Alem\'an \& Trujillo Bueno(2015)]{DelPino2015}
del Pino Alemán, T., \& Trujillo Bueno, J. 2015, The Astrophysical Journal Letters,   808,  L13, 6 pp.  

\bibitem[Del Pino Alem\'an \& Trujillo Bueno(2017)]{DelPino2017}
del Pino Alem\'an, T., \& Trujillo Bueno, J. 2017, ApJ,   838,  article id. 164, 9 pp.
 
\bibitem[Derouich et al.(2005a)]{Derouich2005a}
Derouich, M., Sahal-Br\'echot, S., \& Barklem, P. S. 2005a, A\&A,    434,  779. 

\bibitem[Derouich et al.(2005a)]{Derouich2005b}
Derouich, M., Barklem, P. S., \&  Sahal-Br\'echot, S.  2005b, A\&A,   441, 395.

\bibitem[Derouichetal2007]{Derouichetal2007}  Derouich, M., Trujillo Bueno, J., \& Manso Sainz, R. 2007, A\&A, 472, 269 

\bibitem[Derouich(2020)]{Derouich2020}
Derouich, M. 2020, The Astrophysical Journal Supplement Series,   247,   id.72, 8 pp.   

\bibitem[DerouichQutub(2024)]{DerouichQutub2024}
 Derouich, M., Qutub, S.  2024, A\&A,    683,  id.A173 
 
\bibitem[Fontenlaetal(1990)]{Fontenla1990}
 Fontenla, J. M., Avrett, E. H., \& Loeser, R. 1993, ApJ, 406, 319 

\bibitem[Keller  Sheeley(1999)]{KellerSheeley1999}
 Keller, C. U., \& Sheeley, N. R. Jr. 1999, in Proc. 2nd SPW,
Solar Polarization, ed. K. N. Nagendra, \& J. O. Stenflo,
ASSL (Dordrecht: Kluwer), 243, 17 

 \bibitem[Landi Degl’Innocenti \& Landolfi(1990)]{1990}
      Landi Degl'Innocenti, E.,  Bommier, V., Sahal-Brechot, S.  A\&A, 235,   459.

\bibitem[Landi Degl’Innocenti \& Landolfi(2004)]{LL04}
Landi Degl’Innocenti, E., \& Landolfi, M. 2004, Polarization in Spectral Lines (Dordrecht: Kluwer).

\bibitem[Manso Sainz \& Landi Degl’Innocenti(2002)]{Manso2002}
Manso Sainz, R., \& Landi Degl’Innocenti, E. 2002, A\&A, 394, 1093.

\bibitem[Nienhuis(1976)]{Nienhuis1976}
Nienhuis, G. 1976, J. Phys. B, 9, 167.

\bibitem[Omont(1977)]{Omont1977}
Omont, A. 1977, Prog. Quantum Electron., 5, 69.

\bibitem[Sahal-Bréchot(1977)]{SB1977}
Sahal-Bréchot, S. 1977, ApJ, 213, 887.

\bibitem[Sahal-Bréchot et al.(2007)]{SB2007}
Sahal-Bréchot, S., Derouich, M., Bommier, V., \& Barklem, P. S. 2007, A\&A, 465, 667.

\bibitem[Sheeley Keller (2003)]{Sheeley2003}
 Sheeley, N. R., Jr., \& Keller, C. U. 2003, ApJ, 594, 1085 

\bibitem[Stenflo(1994)]{S1994}
 Stenflo, J., O. 1994, Solar Magnetic Fields: Polarized Radiation Diagnostics, Astrophysics and Space Science Library, Vol. 189. Berlin: Springer, 1994, 385pp
 \bibitem[Trujillo Bueno1999]{Trujillo_1999}
Trujillo Bueno, J. 1999, in Solar Polarization: Proceedings of an International
Workshop Held in Bangalore, India, 12--16 October, 1998, ed.
K. N. Nagendra \& J. O. Stenflo (Astrophysics and Space Science
Library, Vol. 243; Boston, MA: Kluwer Academic), 73 

\bibitem[Trujillo Bueno et al.(2001)]{Trujillo_et_al_2001}
Trujillo Bueno, J.,  Collados, M.,  Paletou, F.,  Molodij, G. 2001, Advanced Solar Polarimetry -- Theory, Observation, and Instrumentation -- 20TH NSO/Sac Summer Workshop, ASP Conference Proceedings, Vol. 236. Edited by Michael Sigwarth. San Francisco: Astronomical Society of the Pacific ISBN: 1-58381-073-0, 2001, p.141.

\bibitem[Trujillo Bueno(2001)]{Trujillo2001}
 Trujillo Bueno, J.  2001
Advanced Solar Polarimetry -- Theory, Observation, and Instrumentation -- 20TH NSO/Sac Summer Workshop, ASP Conference Proceedings, Vol. 236. Edited by Michael Sigwarth. San Francisco: Astronomical Society of the Pacific ISBN: 1-58381-073-0, 2001, p.161 
\end{thebibliography}
\end{document}